\documentclass[11pt]{article}

\usepackage{amsmath,amsfonts, epsfig} 
\usepackage{amssymb}
\usepackage{amscd}
\advance \textheight by \topskip
\makeatletter

\@addtoreset{equation}{section}
 \def\theequation{\thesection.\arabic{equation}}
\makeatother

\newtheorem{theorem}{Theorem}[section]
\newtheorem{definition}[theorem]{Definition}
\newtheorem{corollary}[theorem]{Corollary}

\newtheorem{proposition}[theorem]{Proposition}

\newcommand{\beqa}{\begin{eqnarray}}
\newcommand{\eeqa}{\end{eqnarray}}
\newcommand{\noi}{\noindent}

\begin{document}

\begin{titlepage}
         \begin{center}
 \vskip .5in
 \begin{flushright}  ICMPA-MPA/2008/17 \\
LPT Orsay-08-58\\
IHES/P/08/43
 \end{flushright}

 \begin{center}
 \vskip .5in

 {\LARGE\bf  One-loop $\beta$ functions of a
translation-invariant renormalizable noncommutative scalar model
}
 \end{center}

  \begin{center}
 {\bf Joseph Ben Geloun$^{a,b,c,\dag}$ and Adrian Tanasa$^{d,e, \ddag}$ }\\

\vspace{0.5cm}
$^{a}$International Chair of Mathematical Physics and Applications\\ 
Universit\'e d'Abomey-Calavi\\
ICMPA-UNESCO Chair, 072 B.P. 50, Cotonou, Rep. of Benin\\
$^{b}$Facult\'e des Sciences et Techniques, Dpt. Math\'ematiques-Informatique \\
Universit\'e Cheikh Anta Diop, Dakar, S\'en\'egal\\
$^{c}$Laboratoire de Physique Th\'eorique, UMR CNRS 8627\\
b\^at. 210, Universit\'e Paris-Sud X1, 91405 Orsay, France\\
$^d$Departamentul de Fizica Teoretic\u a,\\
 Institutul de Fizica si Inginerie Nuclear\u a Horia Hulubei,\\
P. O. Box MG-6, 077125 Bucure\c sti-M\u agurele, Rom\^ ania\\
$^e$Institut des Hautes \'Etudes Scientifiques (IH\'ES)\\
Le Bois-Marie, 35 route de Chartres\\
F-91440 Bures-sur-Yvette, France\\
Emails: $^\dag$joseph.bengeloun@cipma.uac.bj, $^\ddag$adrian.tanasa@ens-lyon.org
 \vspace{0.5cm}
    \end{center}
\vspace{7.5pt}
 \today
 \begin{abstract}
Recently, a new type of renormalizable  $\phi^{\star 4}_{4}$ scalar model on the Moyal space 
was proved to be perturbatively renormalizable. It is translation-invariant and introduces in the action 
a  $a/(\theta^2p^2)$ term.  We calculate here the $\beta$ and $\gamma$ functions at one-loop level for this model.
The coupling constant $\beta_\lambda$ function is proved to have the same behavior as the one of the $\phi^4$ 
model on the commutative $\mathbb{R}^4$.  The $\beta_a$ function of the new parameter $a$ is also calculated. Some interpretation of these results are done.
 \end{abstract}

\noindent
\end{center}
Pacs numbers: 02.40.Gh, 11.10.Nx.\\
MSC codes: 76F30,  81T75, 46L65,  53D55.

\noindent Key-words:  Noncommutative quantum  field theory, 
renormalization, $\beta$ function.
 \end{titlepage}

\section{Introduction}
Noncommutative (NC) quantum field theories (NCQFT) \cite{DN}\cite{vince}
is intensively investigated in the recent years \cite{GW2}-\cite{gauge2}.
A first renormalizable $\phi^{\star 4}_4$ model on the Moyal space, the Grosse-Wulkenhaar model (GWm), 
was proposed in \cite{GW2}.  Ever since several proofs of its renormalization were given 
and some of its properties were studied \cite{Orsay}-\cite{dimreg}.
The $\beta$ function of this model was proved to be vanishing at any order in 
perturbation theory  \cite{beta1, beta23, beta}.  
These advances motivate to better scrutinize these NC models. 
Moreover, other renormalizable models have been highlighted.
The $O(N)$ and $U(N)$ GWm were considered with respect to symmetry breaking issues \cite{goldstone}; 
their $\beta$ functions were computed at one-loop in \cite{Geloun:2008ir}.  
Finally, the GWm in a magnetic field  was considered with respect to its parametric representation \cite{param2} and 
its $\beta$ function computations at any order \cite{Geloun:2008zk}.

Nevertheless, the GWm mentioned above loses 
the usual translation invariance of a field theory. 
Furthermore, it does not seem easy to generalize this method to gauge theories: one is lead to theories with non-trivial vacua \cite{gauge}, in which renormalizability is unclear up to now.

In \cite{noi},  
a different type of scalar model was proposed.
This model preserves 
translation invariance and is also proved to be renormalizable 
at all order of perturbation theory \cite{noi}.
These features come from 
a new term in the propagator, of the form $a / (\theta^2p^2)$ and on which relies the ``cure'' of the UV/IR mixing.
Finally, let us also mention that the extension of this mechanism for gauge theories was recently proposed \cite{gauge2}.

In this paper, we consider this NC translation-invariant scalar model 
and compute its one-loop $\beta$ functions for the coupling constant $\lambda$, the mass $m$ and 
the new parameter $a$. We decompose the propagator of the theory as a sum of the usual commutative propagator 
and a NC correction. Different comparisons with the commutative $\phi^4$ model are made. 
The sign of the $\beta_\lambda$ function is proved to be the same as in the commutative theory.

The paper is organized as follows.  The next section
introduces the model and recalls some of the renormalization results of \cite{noi}.
The third section proposes the decomposition mentioned above of the NC propagator. 
This decomposition allows us to calculate the $\gamma$ and $\beta$ functions of the model. The last section extends these results to any loop order; some considerations on the physical relevance of the model are also given.
Finally, some conclusions are drawn.

\section{The model and its renormalization}
\renewcommand{\theequation}{\thesection.\arabic{equation}}
\setcounter{equation}{0}

\subsection{The noncommutative model}

The  action \cite{noi} is
\begin{eqnarray}
S_\theta [\phi]=\int d^4 p \left\{ \frac{1}{2}  p_{\mu} \phi\,  p^\mu \phi \, + \, \frac
12 m^2\,  \phi \, \phi   
\,+\, \frac{1}{2} \, a \, \frac{1}{\theta^2 p^2} \,\phi \, \phi  
\, + \,\frac{\lambda }{4}\, V_\theta \right\},
\label{revolutie}
\end{eqnarray}
where  $a $ is  some dimensionless parameter chosen such that 
 $ a \ge 0 \, $ and $V_\theta$ is the Fourier transform of 
the potential $\frac{\lambda}{4}\phi(x)^{*4}$ in momentum space. 
Note that the $4$ factor above (instead of the usual commutative $4!$ factor)
comes from the fact that the Moyal vertex is invariant only under cyclic permutation of the incoming/outgoing fields. 
However, the comparison with the commutative results will become more difficult.
The propagator writes
\begin{eqnarray}
C(p)=\frac{1}{p^2+m^2+\frac{a}{\theta^2 p^2}} \, .\label{propa}
\end{eqnarray}
Note that the condition on $a$ above ensures the positivity of $C(p)$.
It is worthy to recall that the Moyal vertex can be written in terms of momenta as
\begin{eqnarray}
V_{\theta} = -\frac{\lambda}{4!}\, \delta(p^{1}+p^{2}+p^{3}+p^{4})\,
e^{\frac{i}{2}\sum_{1\leq j<\ell\leq 4} p^j_{\mu}\Theta^{\mu\nu} p^\ell_{\nu} }
\end{eqnarray}

Let us introduce the following terminology.
\begin{definition}
\label{def1}
Let $g$ be the genus, $L$ the number of lines, $F$ the number 
of faces, $B$ the number of faces broken by external legs of a
graph. 
\begin{enumerate}
\item[(i)] A planar graph is a graph such that $g=0$.
\item[(ii)] A nonplanar graph is a graph such that $g>0$.
\item[(iii)] A planar regular graph is a planar graph such that $B=1$.
\item[(iv)] A planar irregular graph is a planar graph such that $B\geq 2$.
\end{enumerate}
\end{definition}

\subsection{Multi-scale analysis. Noncommutative renormalization - locality uplifts to ``Moyality''}

As already stated in \cite{noi}, the primitively divergent graphs of the model \eqref{revolutie} are the 
two- 
and four-point ones. More precisely, one has the following:

\begin{itemize}
\item the planar regular two-point graphs are responsible for the wave function and mass renormalization,

\item the planar regular four-point graphs are responsible for the coupling constant renormalization,

\item the planar irregular two-point graphs are responsible for the renormalization of the parameter $a$.

\end{itemize}

\begin{figure}
    \centering
\includegraphics[width=12cm, height=3.5cm]{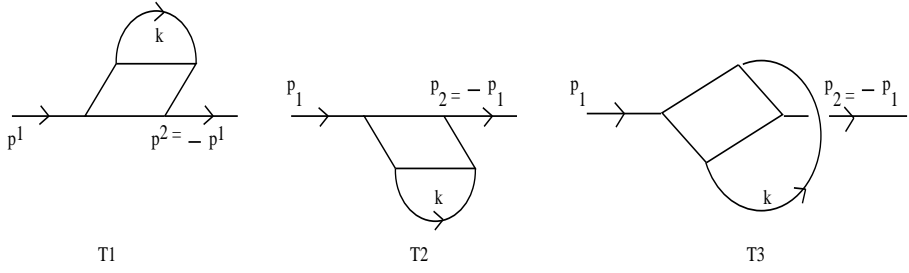}
\caption{ {\small The tadpole graphs $T_1$,  $T_2$ and $T_3$.}}
\label{fig1}
\end{figure}

\begin{figure}
    \centering
\includegraphics[width=6cm, height=2.5cm]{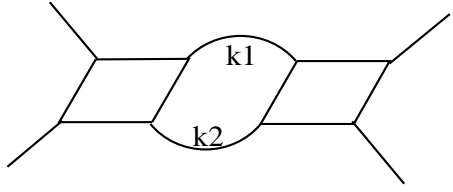}
\caption{ {\small The bubble graph}}
\label{fig2}
\end{figure}

The rest of the graphs are irrelevant to the renormalization process. 
Thus, the one-loop graphs to be considered are the ones of Fig.\ref{fig1} and \ref{fig2}. 
Note that the tadpole graphs $T_1$ and $T_2$ are planar regular graphs while $T_3$ is a planar irregular graph. 

\begin{figure}
    \centering
\includegraphics[width=10cm, height=3cm]{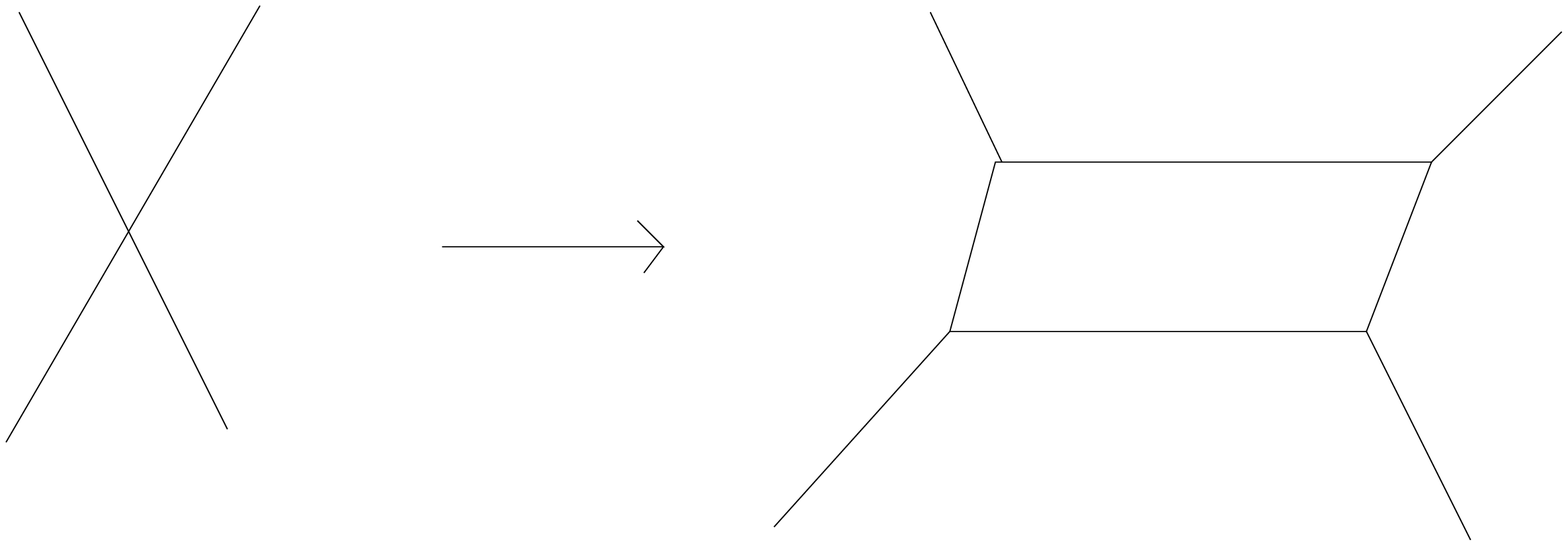}
\caption{ {\small The locality of counterterms is replaced by their ``Moyality''.}}
\label{moyality}
\end{figure}

Before going further let us say a few words about noncommutative renormalization. When going from commutative quantum field theory to NCQFT one loses the principle of position space locality. Nevertheless, what is important in the renormalization process is the fact that the counterterms have the same form as the ones in the original action. In the case of commutative quantum field theory these terms are indeed local. In the case of NCQFT, these terms will no longer be local but they will continue to have the same, ``Moyal form'', as the ones in the original action. Thus one can state that the general principle of locality has been replaced with a new one, of ``Moyality'' (see Fig. \ref{moyality}). Renormalization further continues along the well-known lines (for some discussion on this issue, see for example \cite{vince}. Note that a first proof of this argument, using multi-scale analysis in position space was given for the GWm in \cite{GW4}). Therefore, as for the other properties that we are used to in a renormalization process, one can conclude that the one-loop contributions to the $\beta$ function are scheme independent. Indeed, the difference between the use of various schemes leads only to finite, thus irrelevant, contributions.

Let us also emphasize here that the renormalization proof given in \cite{noi} was based on the multi-scale analysis. We now give a few details of this technique (for a general presentation see \cite{carte}). One cuts the propagator in some appropriated momentum slices, using for practical reasons a geometrical progression of ratio $M$. In momentum space, for a commutative $\phi^4$ theory this writes
\beqa
\frac{1}{p^2+m^2}&=&\sum_{i=0}^\infty C_{\rm comm}^i (p),\nonumber\\
C_{\rm comm}^i (p)&=&\int_{M^{-2i}}^{M^{-2(i-1)}}d\alpha\, e^{-\alpha (p^2+m^2)}\le K_{\rm comm} e^{-c_{\rm comm}M^{-2i(p^2+m^2)}},\ i\ge 1,\nonumber\\
C_{\rm comm}^0 (p)&=&\int_{1}^{\infty}d\alpha\, e^{-\alpha (p^2+m^2)}\le K_{\rm comm} e^{-c_{\rm comm}p^2},
\eeqa
where $K_{\rm comm}$ and $c_{\rm comm}$ are some constants.

One then uses the BPHZ scheme. The renormalization conditions in momentum space write 
\beqa
\label{conditii}
\Gamma^4 (0,0,0,0)=-\lambda_r,\ G^2(0,0)=\frac{1}{m^2},\ \frac{\partial}{\partial p^2}G^2(p,-p)|_{p=0}=-\frac{1}{m^4}.
\eeqa
where 
$\Gamma^4$ and $G^2$ are the connected functions and 
by $r$ we mean renormalized (see again \cite{carte}).

Note that in perturbation theory one defines the high scales as the momentum range where the denominator $p^2+m^2$ of the propagator is large. In the case of the noncommutative theories we do the same, but now the noncommutative propagator \eqref{propa} has now $p^2+m^2+\frac{a}{\theta^2p^2}$ as denominator. One thus has (see \cite{noi})
\beqa
C(p)&=&\sum_{i=0}^\infty C^i (p),\nonumber\\
C^i (p)&=&\int_{M^{-2i}}^{M^{-2(i-1)}}d\alpha\, e^{-\alpha (p^2+m^2+\frac{a}{\theta^2p^2})}\le K e^{-cM^{-2i(p^2+m^2+\frac{a}{\theta^2p^2})}},\ i\ge 1,\nonumber\\
C^0 (p)&=&\int_{1}^{\infty}d\alpha\, e^{-\alpha (p^2+m^2+\frac{a}{\theta^2p^2})}\le K e^{-cp^2},
\eeqa
where again $K$ and $c$ are some constants.
One has thus a different slicing of the propagator  where the UV and IR regions are mixed together. It is actually this new type of slicing which is responsible for the renormalizability of the model.

The renormalization scheme used is again the BPHZ scheme with the same renormalization conditions \eqref{conditii} as in the commutative case, the only difference being that one does not take anymore the value $0$ for the momenta but the minimum $p_m$ of the expression $p^2+\frac{a}{\theta^2p^2}$ (from the reasons explained above). In the case of the $4-$point function one can consider $\Gamma^4(p_m,-p_m,p_m,-p_m)$.

\section{The $\beta$ functions of the model}

\subsection{Decomposition of the propagator: noncommutative correction}

Before calculating the $\beta$ and $\gamma$ functions of this model, 
let us write 
some useful integral representation of the propagator \eqref{propa}.
We propose  to use the formula
\begin{eqnarray}
\frac{1}{A+B}=\frac 1A - \frac 1A B \frac{1}{A+B},
\end{eqnarray}
for 
\begin{eqnarray}
\label{AB}
A=p^2+m^2,\ \ B=\frac{a}{\theta^2 p^2}.
\end{eqnarray}
Thus, the propagator \eqref{propa} writes
\begin{eqnarray}
C(p)&=&\frac{1}{p^2+m^2}-\frac{1}{p^2+m^2}\frac{a}{\theta^2 p^2 (p^2+m^2)+a}\cr
&=&
\frac{1}{p^2+m^2}-\frac{1}{p^2+m^2}\frac{a}{\theta^2 (p^2 +m_1^2)(p^2+m^2_2)},
\label{propa2}
\end{eqnarray}
where $-m_1^2$ and $-m_2^2$ are the roots of the denominator of 
the second term in the lhs considered as a second order equation in $p^2$, namely
\begin{eqnarray}
\frac{-\theta^2 m^2\pm \sqrt{\theta^4 m^4 - 4 \theta^2 a}}{2\theta^2}<0,
\end{eqnarray}
with $a<\theta^2 m^4/4$. 
One can also use the following formula
\begin{eqnarray}
\frac{1}{p^2+m_1^2}\frac{1}{p^2+m_2^2}=
\frac{1}{m_2^2-m_1^2}(\frac{1}{p^2+m_1^2}-\frac{1}{p^2+m_2^2}).
\end{eqnarray}
This allows to write the propagator \eqref{propa2} as
\begin{eqnarray}
\label{propa3}
C(p)=
\frac{1}{p^2+m^2}-\frac{a}{\theta^2 (m_2^2-m_1^2)}
\frac{1}{p^2+m^2}(\frac{1}{p^2+m_1^2}-\frac{1}{p^2+m_2^2}).
\end{eqnarray}
Note that, in this paper, we will use the decomposition \eqref{propa2}, the one given by \eqref{propa3} being equivalent. One can interpret the last term of \eqref{propa2} as some noncommutative correction to the propagator. 
Let us now prove that this correction leads only to irrelevant ({\it i.e.} finite) contribution when inserted into the one-loop diagrams of Fig.\ref{fig1} and \ref{fig2}.

Indeed, when inserted in the $T_1$ or $T_2$ tadpole graphs, we get an integral of the form
\beqa
\lambda \int d^4 p \frac{1}{(p^2+m^2+\frac{a}{\theta^2p^2})}.
\eeqa
Thus the noncommutative correction obtained {\it via} the decomposition \eqref{propa2} is
\beqa
\label{convergenta}
\lambda \int d^4 p \frac{a}{\theta^2}\frac{1}{(p^2+m^2)(p^2+m_1^2)(p^2+m_2^2)},
\eeqa
which is convergent.

The case of the planar irregular tadpole graph $T_3$ induces the same integral 
when letting the external moment go to $0$. Finally, the bubble graph of Fig.\ref{fig2} leads also to a finite integral which is irrelevant. 
This can be explicitly seen by writing the corresponding Feynman amplitude (at vanishing external momenta)
\beqa
\lambda^2 \int d^4 p \frac{1}{(p^2+m^2+\frac{a}{\theta^2p^2})^2}.
\label{lamcar}
\eeqa
Inserting now the decomposition \eqref{propa2} in the integral \eqref{lamcar} implies
the separation of the noncommutative correction
\beqa
\lambda^2 \left[2\frac{a}{\theta^2}\int d^4 p \frac{1}{(p^2+m^2)^2(p^2+m_1^2)(p^2+m_2^2)} + \frac{a^2}{\theta^4}\int d^4p \frac{1}{[(p^2+m^2)(p^2+m_1^2)(p^2+m_2^2)]^2}
\right].
\eeqa
Both these integrals are finite thus irrelevant.

\subsection{One-loop $\beta$ and $\gamma$ functions}

We briefly set the renormalization group (RG) flow framework used hereafter. Firstly,
the dressed propagator $G^2(p)$ or connected two-point function 
is given by 
\begin{eqnarray}
G^{2}(p)&=&\frac{ C(p)}{1-C(p)\Sigma(p)} =\frac{1}{C(p)^{-1} - \Sigma(p)},\\
C(p)^{-1} - \Sigma(p)& =& p^2+m^2+\frac{a}{\theta^2 p^2}- \Sigma(p) 
\end{eqnarray}
where $\Sigma(p)$ is the self-energy.
One writes
$\Sigma(p) = \langle \phi(p)\phi(-p) \rangle^{t}_{1PI}$, 
where by $t$ we understand amputated. 
Furthermore, note that
\begin{eqnarray}
\label{decompozitie}
\Sigma(p) = \Sigma_{\rm plr}(p) + \Sigma_{\rm pli}(p).
\end{eqnarray}
``plr" and ``pli" refer to planar regular and irregular contributions, respectively.

We now want to compute at one-loop the renormalization equations
\begin{eqnarray}
\lambda_{r} = -\frac{\Gamma^{4}(p_m,-p_m,p_m,-p_m)}{Z^2},\qquad
m_{r}^2-m_b^2 = - \frac{\Sigma_{\rm plr}}{Z},
\end{eqnarray}
where by $r$, we mean ``renormalized" and by $b$, we mean ``bare".
In addition, $Z$ is the wave function renormalization and the amputated four-point function is
\begin{eqnarray}
\Gamma^{4}(p^1,p^2,p^3,-p^1-p^2-p^3) =  \langle \phi(p^1)\phi(p^2)\phi(p^3)\phi(-p^1-p^2-p^3) \rangle^{t}_{1PI}
\end{eqnarray}

The RG flow of the parameter $a$ is now considered.
In \cite{noi}, it was already observed that this renormalization is {\it finite}, 
meaning that the coefficient of $1/p^2$ is finite. Indeed, an explicit computation 
of the Feynman amplitude of a planar irregular  two-point function leads to this result (see again \cite{noi})
\beqa
\label{rez}
{\cal A}=\frac{1}{p^2} F(p),
\eeqa
where ${\cal A}$ is the corresponding amplitude and $F(p)$ is a function uniformly bounded by a constant for all $p$.
This is related to the fact that the slice definition takes into consideration the mixing of high and low energies.

One has

\begin{proposition}
\label{lem1}
At one-loop, the self-energy is given by
\begin{eqnarray}
\label{Sigma}
&&\Sigma(p) = - \lambda\,
\left( 2  S^{(1)}(0) 
+S^{(1)}(p)   \right),
\label{eqsig}
\end{eqnarray}
where 
\begin{eqnarray}
S^{(1)}(p) &=&\int d^4 k \frac{e^{i k_\mu \Theta^{\mu \nu} p_\nu}}{k^2+m^2}.
\label{eqs1s22}\\
\end{eqnarray}
\end{proposition}

\noindent{\bf Proof.}
The self-energy can be obtained at first order in $\lambda$ by 
\begin{eqnarray}
\Sigma(p) =\sum_{{\cal G}_i}  K_{{\cal G}_i}S_{{\cal G}_i}(p) 
\label{sigma}
\end{eqnarray}
where ${\cal G}_i$ runs over one-loop 1PI amputated two-point planar regular and irregular graphs
with amplitude $S_{{\cal G}_i}(p)$, 
and $K_{{\cal G}_j}$ corresponds to some combinatorial factor.
As discussed above, the graphs to be considered are the 
tadpole graphs $T_1$, $T_2$ and $T_3$ (see Fig.\ref{fig1}),  
with the combinatorial factors 
\begin{equation}
K_{T_1} = 4,\quad\quad K_{T_2}= 4,\quad\quad
K_{T_3}= 4,
\label{com}
\end{equation}
respectively. 
Since the noncommutative correction of the propagator produces an irrelevant contribution (see above), 
we obtain $S^{(1)}(0)$ for the amplitudes of the tadpole graphs $T_1$ and $T_2$ 
and $S^{(1)}(p)$ for the amplitude of the $T_3$ graph. 
\hfill $\square$

\medskip
Remark that the integral $S^{(1)}(0)$ is quadratically divergent while $S^{(1)}(p)$ is convergent; 
nevertheless it is this integral which leads to the UV/IR mixing 
(indeed, a $1/p^2$ contribution which, when inserting the corresponding planar irregular tadpole 
into a ``bigger'' graph will lead to a IR divergence).

Furthermore, we point out 
that the decomposition \eqref{decompozitie} of the self-energy into a planar regular and a planar irregular part corresponds in \eqref{Sigma} to $2S^{(1)}(0)$ for the planar regular part (the wave function and mass renormalization) and to $S^{(1)}(p)$ (the renormalization of the parameter $a$). Hence, one has a splitting of this self-energy into two distinct parts, responsible for the renormalization of two distinct parameters, $m$ and $a$. This is a major difference with respect to the commutative $\phi^4$ model.

\medskip

Let us calculate the wave function renormalization $Z=1-\partial_{p^2}\Sigma_{\rm plr} (p)|_{p=p_m}$. 
Since $S^{(1)}(0)$ has no dependence on the external momenta $p$, the following one-loop result is reached
\beqa
\label{Z}
Z=1.
\eeqa
Then, the $\gamma$ function of the model is
\beqa
\label{gamma}
\gamma =0+{\cal O}(\lambda^2).
\eeqa
Note that we have proved that the results \eqref{Z} and \eqref{gamma} are at one-loop, 
for the reasons explained above, nothing but the ones of the $\phi^4$ theory on commutative space.

In the following, we  investigate the RG flows of the parameters $m$. As a straightforward consequence of Proposition \ref{lem1}, 
the tadpole graphs $T_1$ and $T_2$ represent $2/3$ of $\Sigma(0)$. The total self-energy at vanishing external momenta 
$\Sigma (0)$  is nothing but the one of the commutative $\phi^4$ model (for a proper rescaling of $\lambda$). We have
\beqa
\beta_m \propto \beta_m^{\rm commutative}.
\eeqa

As a consequence of the above discussion of the {\it finite} renormalization of the parameter $a$ we have
\beqa
\beta_a =0.
\eeqa

We want to emphasize that the splitting \eqref{decompozitie} of the self-energy can also be associated to  
some mechanism for taking the commutative limit, as already indicated in \cite{noi}.

In the following, the RG flow of the coupling constant $\lambda$ is calculated.
The following statement holds.
\begin{theorem}
\label{theo}
At one-loop, the RG flow of the coupling $\lambda$ satisfies
\begin{equation}
\lambda_r = \lambda\;  \left( 1-2\lambda\, 
{\cal S}^{(2)} \right),
\label{lamr}
\end{equation}
with 
\begin{eqnarray}
\label{s2}
{\cal S}^{(2)} = \int d^4 k \frac{1}{(k^2+m^2)^2}.
\label{cals2}
\end{eqnarray}
\end{theorem}

\noindent{\bf Proof.}
 The noncommutative correction of the propagator corresponds to an irrelevant contribution in $\Gamma^4$. 
Only the bubble graph of Fig.\ref{fig2} has to be considered.
Its combinatorial factor is
\begin{eqnarray}
4\cdot 4\cdot 4.
\label{com4}
\end{eqnarray}
The Feynman amplitude of the bubble graph includes the integral $ {\cal S}^{2}$.
Thus, one gets 
\begin{eqnarray}
\Gamma^4(p_m,-p_m,p_m,-p_m)&=& -\lambda +\lambda^2 \frac{1}{2!4^2}
4^3 
  {\cal S}^{2} 
\end{eqnarray}
which completes the proof.
\hfill $\square$

\medskip

Note that the divergence of the integral \eqref{s2} is logarithmic when removing the UV cutoff.
The $\beta_\lambda$ function of the model \eqref{revolutie} is thus a simple fraction of the $\beta_\lambda$ function of the commutative $\phi^4$ model. The difference is due to the fact that one has to take into considerations only the planar regular bubble graph of 
Fig.\ref{fig2}. In other words, the symmetry factor of the noncommutative graph of Fig.\ref{fig2} 
is only a part of the symmetry factor of the corresponding commutative graph
 (for a commutative theory, the planar irregular or regular four-point graphs are indistinguishable). 

With our conventions, after performing the solid angle integration of $d^4k$ (introducing a $2\pi^2$ factor),
one obtains
\beqa
\beta_\lambda={4\pi^2 \lambda^2}{}+{\cal O}(\lambda^3).
\eeqa

\bigskip

Let us argue here about the possibility of level mixing between the renormalization of the kinetic part and the ``a'' part of the propagator. In \cite{noi} it was proved that the $2-$point function with only one broken face leads (at any order in perturbation theory) to
\beqa
\Lambda^2 + p^2 {\rm log}\, \Lambda +\,  {\rm finite\, terms}.
\eeqa
One recognizes here the mass divergence as well as the usual logarithmically divergent wave function renormalization. Furthermore, as already stated above, the $2-$point function with two broken face leads to a behavior (again at any order in perturbation theory) of type \eqref{rez}. We therefore conclude that one has no such level mixing in the model \eqref{revolutie}, result which confirms the conclusion of \cite{noi}.



\section{Any order behavior; physical relevance of the model}

In this final section we investigate the behavior of the RG flow at any order in perturbation theory. We then make some comments on the physical relevance of the model \eqref{revolutie} analyzed in this letter.

\subsection{Any order behavior of the $\beta$ functions}

As already stated above, in \cite{noi} it was proved that the finite character of the renormalization of the parameter $a$ holds at any order in perturbation theory. Thus, one has $\beta_a=0$ at any order.

Furthermore, with respect to the RG flows of the other parameters of \eqref{revolutie}, one has the following result:

\begin{proposition}
The noncommutative correction \eqref{propa2} of the propagator leads to irrelevant contribution for the RG flow at any order in perturbation theory. 
\end{proposition}
{\bf Proof 1.} In order to obtain the largest divergent integral coming from some noncommutative correction \eqref{propa2} of the propagator, one has to consider, amongst the $2^L$ terms of a Feynman amplitude ($L$ being the number of internal lines), the one involving the product of the noncommutative correction of one of the propagators with the commutative parts of the remaining $L-1$ propagators (obviously, the rest of the corrections will be more convergent, since more powers of momenta are added at the denominator). Denoting this particular internal momentum with $k_1$ (for which we consider the noncommutative correction), this term writes
\beqa
\label{conv}
-\frac{a}{\theta^2}\int d^4 k_1 \ldots d^4 k_b \frac{1}{(k_1+m^2)(k_1+m^2_1)(k_1+m^2_2)}
\frac{1}{k_2+m^2}\ldots \frac{1}{k_L+m^2},
\eeqa
where $b$ is the number of loops. Note that since we deal only with planar graphs (see above), one has
\beqa
\label{loops}
b=F-1,
\eeqa
where $F$ is the number of faces of the respective graph. A power counting argument in \eqref{conv} leads to the conclusion that, if one wants this integral to be divergent the condition
\beqa
\label{cond}
b\ge \frac 12 L +1
\eeqa
needs to be fulfilled. From
\beqa
L=2n - \frac 12 N, \ \ \ \ 2=n-L+F
\eeqa
(where $n$ is the number of vertices and $N$ is the number of external legs) one obtains
\beqa
\label{L}
L=2b-2+\frac 12 N
\eeqa
(where \eqref{loops} have been used). Inserting now \eqref{L} in the condition \eqref{cond} leads to a condition which can never be satisfied for $N>0$. Hence the integral \eqref{conv} will be convergent for any loop number $b$.

\medskip

\noi
{\bf Proof 2.} The first largest noncommutative correction adds up a factor $4$ in the power counting of the denominator, with respect to the ``usual'', commutative contribution (as explained above). Nevertheless, the divergences to deal with in this model are logarithmic or quadratic. In both cases, the supplementary factor $4$ in the power counting leads directly, to a convergent, thus irrelevant, contribution in the Feynman amplitude. (QED).

\begin{corollary}
At any order in perturbation theory, the $\beta$ functions of the model \eqref{revolutie} are rational multiples of the corresponding $\beta$ functions of the commutative $\phi^4_4$ model.
\end{corollary}
{\bf Prof.} It is a direct consequence of the proposition above and of the decomposition \eqref{propa2} of the noncommutative propagator.

\subsection{Comments on the physical relevance of the model}

Let us now argue in this subsection on why the noncommutative action \eqref{revolutie} deserves a thorough physical investigation. As already stated in the introduction, the main drawback of the GWm is its translation-invariance breaking, invariance which, in the case of the new type of action \eqref{revolutie}, is now restored. Furthermore, the modification of the propagator with a $\frac{1}{p^2}$ is simply dictated by the complete propagator ({\it i.e.} the propagator dressed with its quantum corrections). Indeed, when computing this for the  ``naive" $\phi^{\star 4}_4$ model, {\it i.e.} the noncommutative scalar model without an harmonic GW $x^2$ or a $1/p^2$ 
term) the $1/p^2$ corrections appears at every order in perturbation theory \cite{limita}.

Let us also state that the increasing interest for the scalar model \eqref{revolutie} has also appeared through recent results, such as, for example, its parametric representation \cite{param-GMRT}. Some one-loop and higher order Feynman diagrams where then explicitly computed, after the release of this preprint, in \cite{high}. Their computation does not use the decomposition \eqref{propa2} of the noncommutative propagator but instead uses explicit Bessel functions forms. Moreover, in \cite{altii}, the static potential associated to the model \eqref{revolutie} was computed.

Furthermore, as also stated in the introduction, when one wants to extend the GW idea to the level of gauge theories, one is lead to theories with highly non-trivial vacua \cite{gauge}. Perturbation theory seems quite cumbersome within this framework. On the other hand, extending the $1/p^2$-term idea to the level of gauge theories leads \cite{gauge2} to an action which preserves this trivial vacua. Work is in progress to investigate the renormalizability of this new type of noncommutative gauge model. 

Let us also add that for the model \eqref{revolutie} a commutative limit mechanism can be written down \cite{limita} (which is not the case so far for the GWm). Indeed, taking into consideration the behavior of the associated Feynman amplitudes (which, as stated above, dictates the appearance of this $1/p^2$-term) one can take the commutative limit by writing in a careful way the counterterms of the model. The main ingredient of this mechanism is that, when $\theta$ is switched to $0$ one just has to write instead of the $1/p^2$ term some mass counterterm (since the convergent integral leading to $1/p^2$ when $\theta\ne 0$ becomes, when $\theta\to 0$, a quadratically divergent integral responsible for the ``usual'' mass and wave function renormalization of the commutative $\phi^4_4$). Such a mechanism could also be useful for the gauge theories of type \cite{gauge2}. 
For a  short review of all these recent developments, one can see \cite{io}.

\bigskip

We have thus computed here the one-loop $\beta$ and $\gamma$ functions of the NC translation-invariant renormalizable scalar model \eqref{revolutie}. 
The $\beta_\lambda$ function is proved to have the same behavior as in the commutative $\phi^4$ case.
This result is a direct consequence of the fact that  $\beta_\lambda$ is given only by the planar regular sector 
of the theory and this sector is not affected by noncommutativity (note that this is the same as 
for the ``naive" NC $\phi^{\star 4}$ model, {\it i.e.} the NC scalar model without an harmonic GW $x^2$ or a $1/p^2$ 
term which presents the UV/IR mixing). 
Indeed, the ``new" planar irregular sector which, in the case of a NC theory, is qualitatively different of the planar regular one,
is responsible only for the renormalization of the constant $a$ (as already observed in \cite{noi}). 
Finally, we have also calculated the running $\beta_a$ of this new constant $a$. 

\section*{Acknowledgements}

J. B. G. thanks  the ANR Program ``GENOPHY" and the  Daniel Iagolnitzer Foundation (France)
for a research grant in the LPT-Orsay, Paris Sud XI.  A. T. thanks the European Science
Foundation Research Networking Program ``Quantum Geometry and Quantum Gravity'' for
the Short Visit Grant 2298.
The authors furthermore acknowledge A. de Goursac and V. Rivasseau 
 for 
 fruitful discussions during the various stages of the preparation of this work. We also thank C. Kopper and J. Magnen for help in proving equations \eqref{propa2} and \eqref{propa3}.

\end{document}